%% file: chiab.tex
\documentclass{aa} 
\usepackage{graphics} 
\input psfig.sty 
\newcommand{\ltsima} {$\; \buildrel < \over \sim \;$} 
\newcommand{\gtsima} {$\; \buildrel > \over \sim \;$} 
\newcommand{\lta} {\lower.5ex\hbox{\ltsima}} 
\newcommand{\gta} {\lower.5ex\hbox{\gtsima}}

\begin{document} 
 
\thesaurus{03(11.01.2; 11.05.1; 11.10.1; 11.14.1)}  
\title{The HST view of the FR~I / FR~II dichotomy 
\thanks{Based on observations with the NASA/ESA 
Hubble Space Telescope, obtained at the Space Telescope Science 
Institute, which is operated by AURA, Inc., under NASA contract NAS 
5-26555 and by STScI grant GO-3594.01-91A}} 
 
\author{M. Chiaberge \inst{1}, A. Capetti \inst{2}  \and 
A. Celotti \inst{1}} 
 
\offprints{M. Chiaberge} 
 
\institute{SISSA/ISAS, Via Beirut 2-4, I-34014 Trieste, Italy (chiab@sissa.it)
\and 
Osservatorio Astronomico di Torino, Strada Osservatorio 20,  I-10025 
Pino Torinese (TO), Italy } 
 
\date{Received 17 June 1999; accepted 21 December 1999} 
 
\titlerunning{HST view of the FR~I and FR~II dichotomy} 
\authorrunning{Chiaberge et al.}   \maketitle 
 
\begin{abstract} 
In order to explore how the FR~I / FR~II dichotomy is related to 
the nuclear properties of radio galaxies, we studied a complete sample 
of 26 nearby FR~II radio galaxies using Hubble Space Telescope 
(HST) images and compared them with a sample of FR~I 
previously analyzed.  FR~I nuclei lie in the radio-optical 
luminosity plane along a tight linear correlation, which argues for a 
common synchrotron origin.  FR~II show a more complex behavior, which 
is however clearly related to their optical spectral classification. 
 
Broad line FR~II radio galaxies (BLRG) are located overall well 
above the FR~I correlation, suggesting that a contribution from thermal 
(disc) emission is present. Three narrow line (NLRG) and one 
weak line radio galaxy (WLRG), in which no nuclear source is 
seen, can be interpreted as the obscured counter-parts of BLRG, in 
agreement with the current unification schemes. 
 
Conversely, in 5 sources of the sample, all of them NLRG or WLRG, 
optical cores are located on the same correlation defined by FR~I and 
with similar radio and optical luminosities.  This suggests that, 
in analogy to FR~I, the emission is dominated by synchrotron 
radiation and represents the optical counter-part of the  
non--thermal radio cores. Interestingly, all these galaxies are 
located in clusters, an environment typical of FR~I. 
 
These results imply that, at least at low redshifts, the FR~II 
population is not homogeneous. Furthermore, the traditional dichotomy 
between edge darkened and brightened radio morphology is not 
univocally connected with the innermost nuclear structure, as we 
find FR~II with FR~I--like nuclei
and this has interesting bearings from the point of 
view of the AGN unified models.
 
\keywords{galaxies: active - galaxies: elliptical and lenticular, cD - galaxies: 
jets - galaxies: nuclei} 
 
\end{abstract} 
 
\section{Introduction} 
\label{intro} 
The original classification of extended radio galaxies by Fanaroff \& 
Riley (\cite{fr}) is based on a morphological criterion, i.e. edge 
darkened (FR~I) vs edge brightened (FR~II) radio structure.  It was 
later discovered that this dichotomy corresponds to a (continuous) 
transition in total radio luminosity (at 178 MHz) which formally 
occurs at $L_{178}= 2\times 10^{33}$ erg s$^{-1}$ Hz$^{-1}$.  The 
presence of radio sources with intermediate morphology, in which 
typical FR~I structures (such as extended plumes and tails) are seen 
together with features characteristics of FR~II sources (narrow jets 
and hot spots) (see e.g. Parma et al.  \cite{parm87}, Capetti et 
al. \cite{capetti95}) argues in favour of a continuity between the two 
classes. 
  
From the optical point of view both FR~I and FR~II are associated with 
various sub-classes of elliptical-like galaxies, but 
statistically their populations are different 
(Zirbel \cite{zirb96}).  Owen (\cite{owen93}) found that 
the FR~I/FR~II division is also linked to the optical magnitude of the 
host galaxy, possibly suggesting that the environment plays an 
important role in producing different extended radio morphologies. 
Moreover, FR~II are generally found in regions of lower galaxy density 
and are more often associated with galaxy interactions with respect to 
FR~I (Prestage \& Peacock \cite{pp}, Zirbel 
\cite{zirb97}). Differences are also observed in the optical 
spectra: while FR~I are generally classified as weak-lined radio 
galaxies, strong (narrow and broad) emission lines are often found in 
FR~II (Morganti et al. \cite{morg92}, Zirbel \& Baum \cite{zirb95}),  
although a sub-class of weak-lined FR~II is also present (Hine \& Longair  
\cite{hine}). 
 
Within the unification scheme for radio-loud AGN (for a review, see 
Urry \& Padovani \cite{urry95}), FR~I and FR~II radio galaxies are 
thought to represent the parent population of BL Lac objects and 
radio-loud quasars, respectively (Antonucci \& Ulvestad \cite{anto85},
Barthel \cite{bart89}).  In order to 
explain the lack of broad lines in the ``mis-oriented'' (narrow-lined) 
FR~II--type objects, obscuration by a thick torus is invoked. A 
combination of obscuration and beaming is therefore necessary at least 
for the FR~II-quasars unification (e.g. Antonucci \& Barvainis \cite{anto90}). 
However, there is evidence that 
this simple picture is probably inadequate: some 
radio--selected BL Lacs - among the most powerful sources in the class 
- display an extended radio structure and luminosity typical of FR~II 
(Kollgaard et al. \cite{koll92}, Murphy et al. \cite{murphy93}) 
and broad - although weak - lines have been observed in some BL Lacs. 
Moreover, Owen et al. \cite{owen96} noted that the  
lack of BL Lacs in a sample of radio galaxies located in Abell clusters 
can be an effect of their selection criteria if the parent population of  
BL Lacs includes both FR~I and FR~II. This idea is also consistent with 
a recently proposed modification of the unification scheme, which claim  
that the weak-lined FR~II are indeed associated with BL Lac objects  
(Jackson \& Wall \cite{jackson}).
These observations can be however reconciled with the unification 
scenario once continuity between the weak and powerful radio--loud 
sources is allowed and thus transition objects are expected. 

In Chiaberge et al. (\cite{pap1}, hereafter Paper~I) we studied HST 
images of all FR~I radio galaxies belonging to the 3CR catalogue, 
finding that unresolved nuclear sources are commonly present in these 
objects. A strong linear correlation is found between this optical and 
the radio core emission, extending over four orders of magnitude in 
luminosity. This, together with spectral information, strongly argues 
for a common non-thermal origin, and suggests that the optical cores 
can be identified with synchrotron radiation produced in a 
relativistic jet, qualitatively supporting the unifying model for FR~I 
and BL Lacs. 
Furthermore, the high rate detection (\gta 85 \%) of optical cores in 
the complete sample indicates that a standard pc--scale geometrically 
thick torus is not present in these low-luminosity radio galaxies. Any 
absorption structure, if present, must be geometrically thin, 
and thus the lack of broad lines in FR~I cannot be attributed to 
obscuration. Alternatively, thick tori are present only in a minority of FR~I.
Given the dominance of non-thermal emission, the optical 
core luminosity also represents a firm upper limit to any thermal 
component, suggesting that accretion might take place in a low 
efficiency radiative regime. 
 
The picture which emerges from this analysis is that FR~Is lack 
substantial thermal (disc) emission, Broad Line Regions and obscuring 
tori, which are usually associated with radio-quiet and powerful 
radio--loud AGN.
 
As a natural extension of Paper~I, here we study the HST images 
of a sample of low redshift FR~II radio galaxies, in order to explore 
how the differences in radio morphology are related to the optical 
nuclear properties. In particular, one of the most important questions 
is whether the FR~I/FR~II dichotomy is generated by two different 
manifestations of the same astrophysical phenomenon, and the 
transition between the two classes is indeed continuous, or instead it 
reflects fundamental differences in the innermost structure of the 
central engine. 
 
The selection of the sample is presented and discussed in 
Sect. \ref{thesample}, while in Sect. \ref{hstobs} we describe the HST 
observations.  In Sect. \ref{CCCII} we focus on the detection and 
photometry of the optical cores. Finally, in Sect. \ref{discussion} we 
discuss our findings. 
 
\section{The sample} 
\label{thesample} 

\begin{figure}
\resizebox{\hsize}{!}{\includegraphics{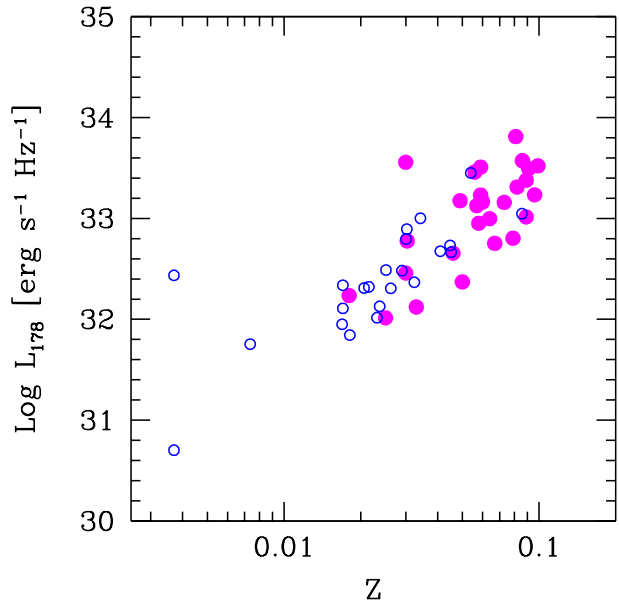}} 
\caption{Total radio luminosity (at 178 MHz) vs redshift diagram for 
the FR~I (empty circles) and FR~II (filled circles) samples.} 
\label{lumz} 
\end{figure}

\input tab1.tex

The sample considered here comprises all radio galaxies belonging to 
the 3CR catalogue (Spinrad \cite{spinrad}) with redshift $z<0.1$, 
morphologically classified as FR~II. We directly checked their 
classification for erroneous or doubtful identifications by 
searching the literature for the most recent radio maps. The final 
list (see Table \ref{tabfr2}) constitutes a complete, flux and 
redshift limited sample of 26 FR~II radio galaxies. 
 
We searched for optical spectral classification and/or optical 
spectra, in order to differentiate our sources on the basis of the 
presence of broad or narrow emission lines.  For only one source 
(namely 3C~136.1) we could not find spectral information in the 
literature.  All spectral types usually associated with FR~II galaxies 
are represented in the sample: five objects are BLRG, fifteen are 
classified as NLRG, while four show only weak lines in their 
optical spectrum (WLRG).  The remaining source, namely \object{3C~371},
has been classified as a BL Lac object.  In Table 
\ref{tabfr2} redshifts and radio data are reported, as taken from the 
literature, together with the optical spectral classifications. 
 
In Fig. \ref{lumz} 
we show the redshift vs total radio luminosity diagram for the sample 
of FR~II galaxies, together with the sample of FR~I discussed in 
Paper~I, but limited to sources with $z<0.1$ for coherence with the 
FR~II sample.  FR~II have a median redshift $z=0.06$, and total radio 
luminosities at 178 MHz are between $10^{32}$ and 10$^{34}$ erg s$^{-1}$ 
Hz$^{-1}$ ($H_0= 75$ km s$^{-1}$ Mpc$^{-1}$ and $q_0=0.5$ are adopted 
hereafter). Notice that whereas the two samples are selected at the 
same limits of redshift and flux, FR~II are, on average, more 
luminous and distant than FR~I.

\section{HST observations} 
\label{hstobs}

HST observations of the FR~II sources are available in the public 
archive (up to April 1999) for 25 out of the 26 sources (only 
\object{3C~33} and \object{3C~105} have not been observed).  The HST 
images were taken using the Wide Field and Planetary Camera 2 
(WFPC2). The whole sample was observed using the F702W 
filter as part of the HST snapshot survey of 3C radio galaxies (Martel 
et al. \cite{martel}, De Koff et al. \cite{De Koff}).  
For 3C~192 we used a F555W image, as this source was not observed with the 
F702W filter. Exposure times are in the range 140--300 s.  The 
data have been processed through the standard PODPS (Post Observation 
Data Processing System) pipeline for bias removal and flat fielding 
(Biretta et al. \cite{biretta}). Individual exposures in each filter 
were combined to remove cosmic rays events. 
 
\section{Optical cores in FR~II} 
\label{CCCII} 
 
\begin{figure*} 
\centerline{\psfig{figure=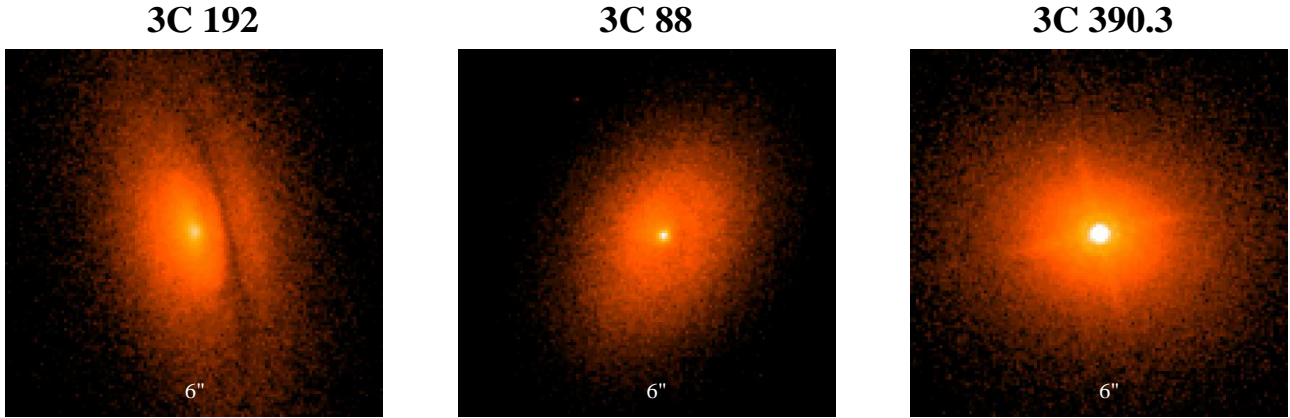,width=0.95\textwidth,angle=0}} 
\caption{HST/WFPC2 broad band images of three FR~II: \object{3C~192} is a NLRG 
showing a diffuse nucleus; \object{3C~388} is a NLRG with an 
optical core; 3C~390.3 is a BLRG.} 
\label{figurine} 
\end{figure*} 

In Paper~I we have adopted a simple operative approach, based  on the
analysis of the nuclear brightness profile, in order to  establish
when an optical core is present in a radio galaxy. 
As in the case of FR~I sources, 
the FWHM fall into two very distinct regimes:
in 11 cases we measured a 
FWHM = 0.05$^{\prime\prime}$ -- 0.08$^{\prime\prime}$,
i.e. indicative of the presence of an unresolved source at the HST
resolution, while in 8 cases we found widths larger than 
0.2$^{\prime\prime}$. We therefore believe that 
no ambiguity exists on whether or not a central unresolved source is
present.

In three sources (3C~382, 3C~390.3 and 3C~445)
the central regions are saturated. While, on the one hand, this 
prevents us from deriving 
their brightness profile, on the other is by itself a clear indication 
of a point-like source. In fact diffuse emission would produce saturation 
with our instrument configuration and exposure times only for surface 
brightness $< 13$ mag arcsec$^{-2}$ in the R band, much larger than 
typically observed
in the central regions of radio-galaxies at this redshift. Furthermore
in all these sources we observe diffraction rings and spikes,
the characteristic hallmarks of the HST Point Spread Function.

We performed aperture photometry of these components. 
The background 
level is evaluated, as in Paper I, by measuring the intensity at a 
distance of $\sim 5$ pixels ($\sim 0.23^{\prime\prime}$) from the center.
The dominant 
photometric error is thus the determination of the background in regions of 
steep brightness gradients, especially for the faintest cores, 
resulting in a typical error of $\sim 10\%$.
For the saturated cores we evaluated their fluxes by comparing the 
PSF wings with those of several bright stars seen in archival HST images
taken with the same filter. This method leads to a somewhat larger 
uncertainties, 25 \% as estimated from the scatter of measures obtained 
with different reference stars, 
which arise from the time dependent structure of the HST PSF.
In Table \ref{tabfr2} we report fluxes and luminosities of 
the optical cores. 

All of the images were taken using broad band filters, which include 
emission lines. In particular, the F702W transmission curve covers the
wavelength range $5900-8200$ \AA ~and thus within our redshift range
includes the H$\alpha$ and [N II] emission lines.
Unfortunately, no HST narrow band images are available for 
the NLRG and WLRG, however, we expect the line contamination 
to be small, due to the wide spectral region  
covered by the filters used ($\sim$ 2000 \AA)
with respect to typical lines equivalent width.  
We correct the broad band fluxes only in the case of BLRG, where the 
emission of broad lines is probably co-spatial to the optical core, 
using ground-based data taken from the literature (Zirbel \& Baum 
\cite{zirbel}). The resulting emission line contribution   
is typically $25- 30 \%$ of the total flux measured in the F702W filter.

In 8 cases the nuclear regions only show diffuse emission: for such 
galaxies we estimate upper limits to a possible central component 
evaluating the light excess of the central 3x3 pixels with respect to 
the surrounding galaxy background.  The remaining 3 sources show 
complex morphologies, e.g. with dust lanes obscuring the central 
regions, and no photometry was performed. 
 
\begin{figure}[h] 
\resizebox{\hsize}{!}{\includegraphics{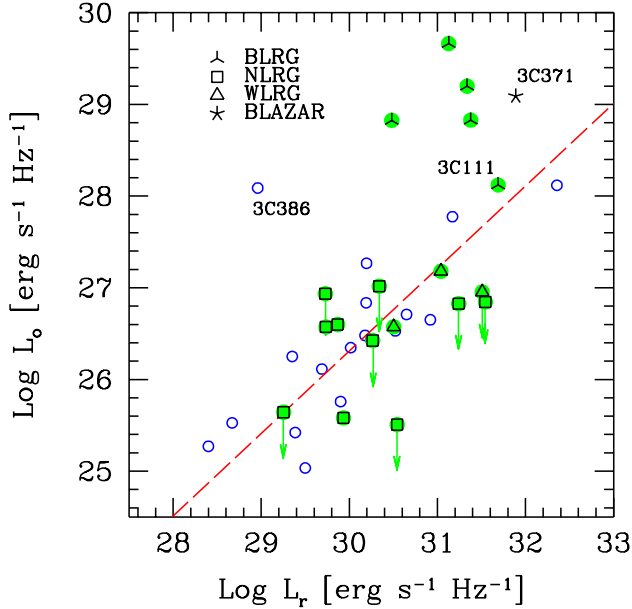}} 
\caption{Optical nuclear luminosity versus radio (5 GHz) core 
luminosity for both the FR~I (open circles) and FR~II (filled circles) 
samples. Different symbols are used to identify different spectral 
classifications. The dashed line is the correlation found in the case 
of FR~I galaxies (see Paper~I).} 
\label{lum}  
\end{figure}

\medskip 
FR~II cores span a wide range of optical luminosities $L_o$ 
(from $10^{25.5}$ up to $10^{30}$ erg s$^{-1}$ Hz$^{-1}$).  In 
Fig. \ref{lum} we report the optical core versus radio (5 GHz) core 
luminosity for the FR~II sample, superimposed to the data (as from 
Paper~I) for FR~I galaxies limiting ourselves, for consistency, to 
those with redshift $z<0.1$.  
FR~II show a complex behavior, which however seems to 
be related to their optical spectral classification. 
 
Let us firstly consider the blazar, \object{3C~371}.  
As one might expect, since its emission is 
dominated by beamed synchrotron radiation, it is among the 
brightest source both in the radio and in the optical band (see 
Fig. \ref{lum}).  In the $L_r$ vs $L_o$ plane it falls in the low 
luminosity end of the region defined by radio selected blazars 
(Chiaberge et al., in preparation).
 
The second group of sources is represented by the BLRG 
(\object{3C~111}, \object{3C~227}, \object{3C~382}, \object{3C~390.3} 
and \object{3C~445}): all of them have very luminous optical cores 
($L_o > 10^{28}$ erg s$^{-1}$ Hz$^{-1}$) being -- together with 
\object{3C~371} -- the most powerful objects of the sample and clearly 
separating from the other FR~II in the diagram. Notice that they 
have an optical excess (or radio deficiency) of up to 2 orders of 
magnitude with respect to the radio-optical core luminosity 
correlation found for FR~I. 
 
In 5 WLRG and NLRG, namely \object{3C~88}, \object{3C~285}, 
\object{3C~388}, \object{3C~402} and \object{3C~403}, we detected 
optical cores which share the same region in the luminosity plane as 
FR~I sources with luminosities between $L_o=10^{25.5}$ up to 
more than $10^{27.5}$ erg s$^{-1}$ Hz$^{-1}$. 
 
The 8 upper limits, all associated with WLRG or NLRG, are also 
plotted. Four objects (\object{3C~35}, \object{3C~98}, 
\object{3C~192} and \object{3C~326}) lie close to the correlation 
defined by FR~I. Conversely, \object{3C~15}, \object{3C~353}, 
\object{3C~452} and \object{3C~236} present an optical luminosity 
deficit of one to two orders of magnitude given their radio core 
emission with respect to FR~I. 
 
No radio core data have been found in the literature for 
\object{3C~136.1}, \object{3C~198} and \object{3C~318.1}. 

\section{Discussion} 
\label{discussion} 
 
In the 24 nearby FR~II 
radio galaxies (out of a complete sample of 26) studied in this paper, 
13 show an unresolved optical core, in 8 cases we can only set an 
upper limit to their luminosity, while 3 sources present a 
complex nuclear morphology.  The location in the optical-radio core 
luminosity plane is clearly connected to the optical spectral 
classification. 
 
Conversely, cores in FR~I radio galaxies show a linear correlation 
between their radio and optical luminosity, which strongly argues for 
a common non-thermal origin.  The presence of such correlation 
provides a useful benchmark to investigate the origin of optical cores 
in FR~II. 
 
In the following we consider each FR~II group separately. 
 
\subsection{Broad Line Radio Galaxies} 
 
Let us first focus on BLRG. These sources present, overall, 
a strong optical excess (or radio deficit), of up to two orders of 
magnitude, with respect to the correlation defined by the FR~I cores 
(see Fig. \ref{lum}).  Note that in the sample of nearby FR~I, the 
only source lying well above the radio-optical correlation is 3C~386, 
which also shows a broad H$\alpha$ line. 
 
BLRG are objects in which the innermost nuclear regions are thought to 
be unobscured along our line of sight (Barthel \cite{bart89}).  We 
therefore expect the presence of a thermal/disc component: 
indeed this might dominate over any synchrotron jet radiation and thus 
be responsible for the observed emission.  The idea that we see 
directly an accretion disc is supported by several observations: 
in the case of \object{3C~390.3}, a bump in the spectral energy 
distribution has been interpreted as radiation emitted by a disc 
component with intermediate inclination (Edelson \& Malkan 
\cite{edelson86}).  Furthermore, the broad and double peaked H$\beta$ 
line observed in this source as well as in other BLRG (see 
e.g. Eracleous \& Halpern \cite{eracl94}), can be accounted for 
within a relativistic accretion disc model (Perez et al. 
\cite{perez88}). 
 
The location of 3C~111 is puzzling, as it lies along the correlation. 
However, several pieces of evidence point to the idea that beamed 
radiation from the relativistic jet significantly contributes in this 
source: it has the largest core dominance among the BLRG of our 
sample; superluminal motions with apparent speed $v \sim 3.4c$ \ have 
been revealed in the inner jet (Vermeulen \& Cohen \cite{vermu}), 
implying that the angle between the line of sight and the jet axis is 
smaller than $\sim 30^{\circ}$; the radio core is strongly variable 
and polarized (Leahy et al. \cite{leahy}).  Furthermore a broad 
K$\alpha$ iron line is detected in the X--ray band, but with a 
relatively small equivalent width which can be explained if the 
continuum emission is diluted by a beamed component (Reynolds et 
al. \cite{reynolds}).  However, its total radio extent of $\sim$ 250 
kpc argues against a viewing angle typical of blazars.  Thus 
\object{3C~111} appears to be a transition source between 
radio-galaxies and blazars, seen at an angle sufficiently small that 
the jet beaming already affects its nuclear properties. 
 
\subsection{WLRG and NLRG with optical cores} 
 
Let us now concentrate on the NLRG and WLRG in which we detected 
optical cores. These objects (2 WLRG and 3 NLRG), all with FR~II radio 
morphology, have cores with radio and optical emission properties that 
are {\it completely consistent with those found in FR~I}.  This 
suggests that in these sources the nuclear emission is   
similarly dominated by synchrotron radiation from the inner jet. 
 
In the case of FR~I, based on the high fraction of detected nuclei, we 
suggested that any obscuring material must be geometrically thin 
and thus the absence of broad lines and the 
relative weakness of any thermal (disc) component with respect to the 
synchrotron emission cannot be ascribed to extinction. 
 
For FR~II with FR~I--like nuclei -- of which we do not have 
enough statistics -- there is an alternative possibility, namely 
that the optical core (jet) emission is produced outside the obscuring 
torus (and thus outside the BLR). In this sense they would represent 
transition objects seen at an intermediate angle between the 
completely obscured and unobscured ones. 
However, VLBI observations 
show that radio cores are unresolved on scale of $\sim$ 0.1 pc 
in nearby radio galaxies and this suggests that their optical 
counter--parts have a similar extent, as already discussed in Paper~I. 
Furthermore, 
a symmetric jet-counter jet structure has been observed in several 
radio sources, implying that they lie essentially in the 
plane of the sky (Giovannini et al. \cite{giovannini98}). If the core 
emission is indeed produced outside the torus, at a distance of, say, 
$\sim$1 pc from the central black hole, a clear separation 
between the two sides of the jet (and no stationary core) should be 
observed in these highly misoriented objects 
(although, at present, symmetric jets have been found only in FR~I). 
This ad hoc geometrical model 
does not seem to be viable, but a conclusive test requires  
spectropolarimetry looking for polarized scattered broad lines. 
A further indication could be obtained from the comparison
of the nuclear infrared (reprocessed?) luminosity of FR~Is and FR~IIs with
FR~I--like nuclei of similar optical luminosity.
 
We conclude that these FR~II are intrinsically narrow-lined objects 
which are in every aspect, except their extended radio-morphology, 
similar to FR~I. Note that the presence of an FR~I-like nucleus 
in a FR~II does not seem to be connected with the total (radio) 
luminosity or redshift of the galaxy. In fact, the total power of such 
sources spans the range $L_{178}= 10^{32}-10^{33.5}$ erg s$^{-1}$ 
Hz$^{-1}$ and the redshifts are between $z=0.025-0.091$, completely 
overlapping with the entire sample and not limited, as one might 
expect, to the low luminosity end. 
 
Conversely, it appears that a possible relationship exists between the 
occurrence of FR~I-like nuclei in FR~II and the environment, as all 
these 5 galaxies reside in clusters. This result can be particularly 
important, as it is known that FR~I and FR~II inhabit different 
environments, with FR~II generally avoiding rich groups, especially at 
low redshifts (Zirbel \cite{zirb97}), while FR~I are usually located 
in rich clusters.  However, before any firm conclusion can be drawn 
about this issue, a larger sample of objects has to be considered. 
 
\subsection{WLRG and NLRG without optical cores} 
 
In 8 galaxies, all WLRG or NLRG, we do not detect the presence of an 
unresolved nuclear component.  
Four sources are located above or very close to the FR~I 
correlation.  They are consistent with being objects in which an 
optical counter--part to the radio core is present,  
but it is too faint 
to be seen against the bright background of the host galaxy. 
 
The remaining 4 objects (3 NLRG and 1 WLRG) are certainly more 
interesting, since they lie 1 - 2 orders of magnitude below the 
correlation. Therefore they lack not only of a BLR, but also of the 
expected optical counterpart of the radio core.  
However, note that these sources have radio core luminosities which cover 
the same range of BLRG.  According to the prescriptions of the 
unification schemes, they can well be the obscured 
counter--parts of BLRG.  Noticeably, excluding the blazar
\object{3C~371}, BLRG and these obscured sources clearly 
distinguish themselves for having the brightest radio cores among FR~II. 
 
\section{Conclusions}

In Chiaberge et al. (\cite{pap1}) we discovered that FR~I nuclei lie 
in the radio-optical luminosity plane along a tight linear 
correlation.  We argued that this is due to a common synchrotron 
origin for both the radio and optical emission. FR~I nuclei must 
also be unobscured and intrinsically lacking of BLR and of 
significant thermal emission from any powerful accretion disc. 

In order to explore how the differences in radio morphology are 
related to the optical nuclear properties, we analyzed HST images of 
24 extended radio-galaxies morphologically classified as FR~II, 
belonging to the 3C catalog and with $z<0.1$.  
We detected optical cores in 13 sources, which implies that the
covering fraction of any obscuring material is less than $\sim 0.54$, or
equivalently, the torus has an opening angle of $\sim 63^{\circ}$.
This can be even larger if at least some of the upper limits
are actually just below the detection threshold. 
Notice that our determination of 
this critical angle is inconsistent with the division
between higher redshift ($0.5 < z < 1$) 3CR quasars and radio galaxies, which
has been found to be $\theta \sim 45^{\circ}$ (Barthel \cite{bart89}).
This might be a problem, however the low redshift selection of our
sources does not allow to derive any firm conclusion. We are currently 
studying a larger and higher redshift sample in order to further investigate 
this issue (Chiaberge et al. in preparation).

Our results suggest that the radio morphology is not univocally connected with 
the optical properties of the innermost structure of radio 
galaxies.  In fact, at least at low redshifts, there is not a single 
homogeneous population of FR~II: unlike FR~I, they show a complex 
behavior, which is however clearly related to their optical spectral 
classification. 
 
In BLRG optical nuclei are likely to be dominated by thermal (disc) 
emission. As discussed above, line emission contamination
cannot account for this excess.
In agreement with the current unification scheme of radio 
loud AGNs, we also identify their possible obscured counter--parts. 
It seems that broad lines and obscuring tori are closely linked and 
both are present only associated to radiatively efficient accretion. 
 
We also find five FR~II sources, spectrally 
identified as narrow lined objects, which harbor nuclei 
essentially indistinguishable from those seen in FR~I. 
By analogy with FR~I, we argue that their optical nuclear
emission is produced primarily by synchrotron radiation,
they are not obscured to our line of sight and therefore 
intrinsically lack a BLR.

Clearly, a classification based on the optical nuclear 
properties, as seen in these HST images, is more likely to reflect 
true similarities (or differences) on the nature of the central 
engine (such as, e.g., the rate of radiative dissipation in the 
accretion disc) than the traditional dichotomy of radio morphology. 
 
From our data and within the limits of the available statistics,  
we find no evidence of a continuous transition between 
the two classes (FR~I and FR~II),
as they are well separated in the $L_r$ vs $L_o$ 
plane. At this stage we only point out that sources with cores below 
$L_o < 10^{27.5}$ erg s$^{-1}$ Hz$^{-1}$ (or equivalently $L_r < 
10^{31}$ erg s$^{-1}$ Hz$^{-1}$) have FR~I like nuclei, while FR~II 
start above this threshold. 
 
It is of particular interest that a significant fraction of FR~II 
(at least 30 \%, but can be as large as 50 \% depending on the nature 
of the sources without detected optical nuclei) have FR~I--like 
nuclei.  The fact that all of these are located in clusters, an 
environment typical of FR~I, might represent an important hint on the 
origin of the different flavours of radio galaxies, worth exploring 
through the study of a larger sample of objects. 
 
These results have also interesting bearings from the point of 
view of the unified models. In fact, this picture argues 
against the idea that all FR~II radio galaxies constitute the parent 
population of radio-loud quasars. We propose instead that 
galaxies with FR~II morphology and an FR~I-like core are possibly 
mis-aligned counter--parts of BL Lac objects.  
This can account for the observation that some 
radio-selected-type BL Lacs show radio morphologies more consistent 
with FR~II than with FR~I (e.g. Kollgaard et al. \cite{koll92}). 
 
To conclude, we note that all of the galaxies included in our sample 
are low redshift objects with total radio powers not exceeding 
$L_{178} \sim 10^{27}$ erg s$^{-1}$ Hz$^{-1}$: thus a crucial 
observational issue is to understand whether these results hold 
to higher power/redshift samples or they are limited to low 
luminosities FR~II. This will be explored in a forthcoming paper. 
 
\begin{acknowledgements} 
We thank Jim Pringle for insightful suggestions and Edo Trussoni 
for useful comments on the manuscript.
The authors acknowledge the Italian MURST for financial support.
This research was supported in part by the 
National Science Foundation under Grant No. PHY94-07194 (A. Celotti).\\ 
This research has made use of the NASA/IPAC Extragalactic Database 
(NED) which is operated by the Jet Propulsion Laboratory, California 
Institute of Technology, under contract with the National Aeronautics 
and Space Administration.\\  
\end{acknowledgements}

\end{document}

%% file: tab1.tex
\begin{table*} 
\caption{Summary of FR~II radio and optical data.}
\hspace{0.5cm} 
\begin{tabular}{l c c c c c c} \hline

Name   &  Redshift &  Spectral   & Log $L_{178}$ & Log $L_r$ (5GHz)  &   $F_o$ (7000 \AA) & Log $L_o$ (7000 \AA)\\
 &     z  & class.  & erg s$^{-1}$ Hz$^{-1}$ & erg s$^{-1}$ Hz$^{-1}$ & erg s$^{-1}$ cm$^{-2}$ Hz$^{-1}$ & erg s$^{-1}$ Hz $^{-1}$\\  
\hline
									          
   3C15    &    0.073   &     NLRG   &      33.16     &   31.54     &   $ <$--28.12   &  $ <$26.85       \\
   3C33    &    0.059   &     NLRG   &      33.51     & 30.27  & \multicolumn{2}{c}{no HST observations} \\
   3C35    &    0.067   &     NLRG   &      32.75     &   30.27     &   $ <$--28.47   &  $ <$26.42      \\
   3C40    &    0.018   &     WLRG   &      32.24     &   30.60     &   $  $   --    &   $  $   --      \\
   3C88    &    0.030   &     WLRG   &      32.46     &   30.50     &   $  $~~--27.64 &  $  $~~26.57      \\
   3C98    &    0.030   &     NLRG   &      32.78     &   29.25     &   $ <$--28.58   &  $ <$25.64      \\
   3C105   &    0.089   &     NLRG   &      33.38     & 30.36  & \multicolumn{2}{c}{no HST observations}  \\
   3C111   &    0.049   &     BLRG   &      33.18     &   31.69     &   $  $~~--26.51 &  $  $~~28.12      \\
   3C136.1 &    0.064   &        -   &      33.00     &    --	    &   $  $   --    &   $  $   --      \\
   3C192   &    0.060   &     NLRG   &      33.16     &   29.73     &   $ <$--27.86   &  $ <$26.94      \\
   3C198   &    0.082   &     NLRG   &      33.31     &    --	    &   $  $~~--27.10 &  $  $~~27.96      \\
   3C227   &    0.086   &     BLRG   &      33.57     &   30.48     &   $  $~~--26.27 &  $  $~~28.83      \\
   3C236   &    0.099   &     WLRG   &      33.52     &   31.51     &   $ <$--28.26   &  $ <$26.95      \\
   3C285   &    0.079   &     NLRG   &      32.80     &   29.93     &   $  $~~--29.44 &  $  $~~25.58      \\
   3C318.1 &    0.046   &     NLRG   &      32.66     &    --	    &   $  $~~--28.94 &  $  $~~25.63      \\
   3C321   &    0.096   &     NLRG   &      33.23     &   30.78     &   $  $   --    &   $  $   --      \\
   3C326   &    0.089   &     NLRG   &      33.02     &   30.34     &   $ <$--28.11   &  $ <$27.02      \\
   3C353   &    0.030   &     NLRG   &      33.56     &   30.54     &   $ <$--28.71   &  $ <$25.51      \\
   3C371   &    0.050   &    BL Lac  &      32.37     &   31.89     &   $  $~~--25.56 &  $  $~~29.09      \\
   3C382$^{*}$   &    0.058   &     BLRG   &      32.95     &   31.13     &   $  $~~--25.11$^{*}$ &  $  $~~29.66$^{*}$      \\
   3C388   &    0.091   &     WLRG   &      33.49     &   31.04     &   $  $~~--27.96 &  $  $~~27.18      \\
   3C390.3$^{*}$ &    0.056   &     BLRG   &      33.46     &   31.38     &   $  $~~--25.91$^{*}$ &  $  $~~28.83$^{*}$      \\
   3C402   &    0.025   &     NLRG   &      32.01     &   29.73     &   $  $~~--27.49 &  $  $~~26.57      \\
   3C403   &    0.059   &     NLRG   &      33.23     &   29.87     &   $  $~~--28.19 &  $  $~~26.60      \\
   3C445$^{*}$   &    0.057   &     BLRG   &      33.13     &   31.34     &   $  $~~--25.56$^{*}$ &  $  $~~29.20$^{*}$      \\
   3C452   &    0.081   &     NLRG   &      33.81     &   31.24     &   $ <$--28.22   &  $ <$26.83      \\

\hline

\end{tabular}
\label{tabfr2}

\medskip
$L_{178}$ and $L_r$ are the total (at 178 MHz) and core radio luminosities, 
taken from the literature. $F_o$ is the flux of the optical core.
Total radio luminosities are calculated from K\"uhr et al. (\cite{kuhr}) or Gower et al. 
(\cite{gower}). Radio core luminosities are taken from Zirbel \& Baum (\cite{zirbel}).
``$<$'' indicate upper limits, i.e. not detected optical cores, ``*'' 
indicate objects in which the optical core saturates, ``--'' in the $F_o$ and $L_o$ colums
mark objects with complex nuclear morphologies for which we do not estimate the core flux and
 ``--'' in the $L_r$ column indicate unavailable radio core data in the literature.

\end{table*}